\documentstyle[preprint,aps,prl,epsf]{revtex}
\tightenlines
\begin{document}

\draft
\title{\hfill MZ-TH/98-14\\[5mm]
One-Pion Charm Baryon Transitions \\
in a Relativistic Three-Quark Model}
\author{M.\ A.\ Ivanov}
\address{Bogoliubov Laboratory of Theoretical Physics, Joint Institute
for Nuclear Research, 141980 Dubna, Russia}
\author{J.\ G.\ K\"{o}rner}
\address{Johannes Gutenberg-Universit\"{a}t, Institut f\"{u}r Physik,
Staudinger Weg 7, D-55099 Mainz, Germany}
\author{V.\ E.\ Lyubovitskij}
\address{Bogoliubov Laboratory of Theoretical Physics, Joint Institute
for Nuclear Research, 141980 Dubna, Russia
and Department of Physics, Tomsk State University,
634050 Tomsk, Russia}
\author{A.\ G.\ Rusetsky}
\address{Bogoliubov Laboratory of Theoretical Physics, Joint Institute
for Nuclear Research, 141980 Dubna, Russia
and HEPI, Tbilisi State University, 380086 Tbilisi, Georgia}
\date{\today}
\maketitle
\begin{abstract}\widetext
We study one-pion transitions between charm baryon states in the
framework of a relativistic three-quark model. We calculate the
charm baryon-pion coupling factors that govern the S-wave, P-wave and
D-wave one-pion transitions from the $s$-wave and the lowest lying $p$-wave
charm baryon states down
to the $s$-wave charm baryon states. For these we obtain:
$g_{\Sigma_c\Lambda_c\pi}$=8.88 GeV$^{-1}$,
$f_{\Lambda_{c1}\Sigma_c\pi}$=0.52 and
$f_{\Lambda_{c1}^*\Sigma_c\pi}$=21.5 GeV$^{-2}$. We compare our rate
predictions for
the one-pion transitions with experimental results.
\end{abstract}

\vspace*{.5cm}
The last few years have seen significant progress in the study of the
spectroscopy
of ground state and excited state charm baryons and their strong one-pion
decays~\cite{PDG}-\cite{E687}. The CLEO~\cite{CLEO1}-\cite{CLEO4},
ARGUS~\cite{ARGUS}, and E687~\cite{E687} Collaborations
reported on measurements of the mass difference between charm baryon
states~\cite{CLEO1}-\cite{E687}, upper limits for their
total widths~\cite{PDG,CLEO2,CLEO3} and first estimations of excited
$\Sigma_c^*$ baryon decay rates~\cite{CLEO4}. Due to the small release
of energy in these transitions the analysis of the
one-pion decays of charmed baryons provide an excellent laboratory for
tests of heavy quark symmetry predictions on the one hand and tests of
the soft dynamics of the light-side one-pion diquark transitions on the
other hand.
The one-pion transitions among charm baryons have been analyzed before
in the framework of the Heavy Hadron Chiral Perturbation Theory
 (HHCHPT)~\cite{Pirjol,Yan,Cheng,Chiladze}, in the
Constituent Quark Model~\cite{Hussain,Pirjol} and in the
Light-Front Quark Model~\cite{Tawfiq}. In the HHCHPT analysis one makes no
assumptions about the composition of the light-side states involved in
the one-pion transitions of the heavy baryons. Using the measured rates of
$\Sigma_c^{*}$ baryons
$\Gamma(\Sigma_c^{*0}\to\Lambda_c\pi^-)$=13.0$^{+3.7}_{-3.0}$ MeV and
$\Gamma(\Sigma_c^{*++}\to\Lambda_c\pi^+)$=17.9$^{+3.8}_{-3.2}$ MeV one
can give estimates of the
unknown coupling parameters appearing in the effective chiral Lagrangian
~\cite{Pirjol,Yan,Cheng,Chiladze}. In
the Constituent Quark Model and Light-Front Quark Model approaches one
further assumes that the light-side state is composed of two constituent quarks.
Using one more assumption still, namely that the one-pion transitions are
governed by
single quark transitions the authors of~\cite{Hussain,Pirjol} were able to
derive a number of relations between the various one-pion coupling constants
of the charm baryons.
The Constituent Quark Model calculations of~\cite{Hussain,Pirjol} did not
address the full dynamics issue in as far as no attempt was made to model
and to calculate the complete wave function overlap of the states involved
in the transition. A first true dynamical calculation of the one-pion
couplings charm baryons characterizing the S-wave, P-wave and D-wave
transitions $g_{\Sigma_c\Lambda_c\pi}$, $f_{\Lambda_{c1}\Sigma_c\pi}$ and
$f_{\Lambda_{c1}^*\Sigma_c\pi}$ was done in~\cite{Tawfiq} where use was made
Light-Front quark model spin functions \cite{Tawfiq}.

In this paper we report on the predictions of the Relativistic Three-Quark
Model~\cite{YadFiz,ZPhys} for the one-pion transitions between
charm baryon states. As in the Light-Front model~\cite{Tawfiq} the
Relativistic Three-Quark Model allows for a full dynamical evaluation
of the one-pion transition strengths between charm baryons.
We want to mention that the Relativistic Three-Quark Model approach was
successfully applied
before to a number of dynamical problems involving the properties of
pions~\cite{YadFiz,ZPhys,PIGG},
light baryons~\cite{FBS} and heavy-light baryons~\cite{PRD97,PRD98,MPL98}.

The Lagrangian describing the couplings of a heavy baryon state to its
constituent light and heavy quarks considerably simplifies in the heavy quark
limit. One has
\begin{eqnarray}
{\cal L}_{B_Q}^{\rm int}(x)&=&g_{B_Q}\bar B_Q(x)
\Gamma_1 Q^a(x)\int\hspace*{-0.1cm} d\xi_1\hspace*{-0.1cm}
\int\hspace*{-0.1cm} d\xi_2
F_B(\xi_1^2+\xi_2^2)\\ \label{int_bar}
&\times&q^b(x+3\xi_1-\sqrt{3}\xi_2)C\Gamma_2\lambda_{B_Q}
q^c(x+3\xi_1+\sqrt{3}\xi_2)
\varepsilon^{abc}+{\rm h.c.}
\nonumber\\
&&\nonumber\\
F_B(\xi_1^2+\xi_2^2)&=&
\int\hspace*{-0.1cm}\frac{d^4k_1}{(2\pi)^4} \hspace*{-0.1cm}
\int\hspace*{-0.1cm}\frac{d^4k_2}{(2\pi)^4} \hspace*{0.1cm}
e^{ik_1\xi_1+ik_2\xi_2}
\tilde F_B\biggl\{\frac{[k_1^2+k_2^2]}{\Lambda_B^2}\biggr\}
\nonumber\\
&&\nonumber\\
&&\nonumber\\
{\cal L}_\pi^{\rm int}(x)&=&\frac{ig_\pi}{\sqrt{2}}\vec\pi(x)
\int\hspace*{-0.1cm} d\xi
F_\pi(\xi^2)\bar q(x+\xi/2)\gamma^5\vec\lambda_\pi q(x-\xi/2)
\\   \label{int_pi}
&&\nonumber\\
F_\pi(\xi^2)&=&
\int\hspace*{-0.1cm}\frac{d^4k}{(2\pi)^4}
e^{ik\xi} \tilde F_\pi\biggl\{\frac{k^2}{\Lambda_\pi^2}\biggr\}
\nonumber
\end{eqnarray}
where $\Gamma_i$ and $\lambda_{B_Q}$ are spin and flavor matrices,
respectively; $g_{B_Q}$ and $g_\pi$ denote the couplings of the respective
hadrons with
the constituent quarks; $\Lambda_B$ ($\Lambda_\pi$) are the cutoff
parameters defining the distributions of light quarks in the heavy baryon
(pion). The baryon cutoff parameter $\Lambda_B$ is chosen to be
the same for charm and bottom baryons such that one has the correct
normalization of the baryonic Isgur-Wise function \cite{PRD97} in the heavy
quark symmetry limit.
The quantum numbers and matrices $\Gamma_i$ and $\lambda_{B_Q}$ define the
structure of the relevant three-quark charm baryon currents and are
listed in TABLE I. The square
brackets $[...]$ and curly brackets $\{...\}$ denote antisymmetric and
symmetric flavour and spin combinations of the light degrees of freedom.

The vertex function which defines the matrix element of the process
$B^i_Q(p)\to B^f_Q(p^\prime)+\pi(q)$ (see Fig. I)
has the following form in the heavy quark limit
\begin{eqnarray}\label{vertex}
M_{inv}^\pi(B_Q^i\to B_Q^f \pi) =
\frac{g_\pi}{\sqrt{2}} g^i_{\rm eff} g^f_{\rm eff}
C_{\rm flavor}\cdot \bar u(v)\Gamma_1^f \frac{(1+\not\! v)}{2}\Gamma_1^iu(v)
\cdot I_{q_1q_2}^{if}(v,q)
\end{eqnarray}
$$
g^2_{\rm eff}=\frac{4C_{\rm color}}{(4\pi)^4} \Lambda_B^4 g_{B_Q}^2,
\hspace{.5cm}
C_{\rm flavor}={\rm tr}\biggl(\lambda_\pi(\lambda_{B^i}+\lambda_{B^i}^\dagger)
(\lambda_{B^f}-\lambda_{B^f}^\dagger)\biggr), \hspace{.5cm} C_{\rm color}=6
$$
\begin{eqnarray}\label{int}
I_{q_1q_2}^{if}(v,q)&=&
\int\hspace*{-0.1cm}\frac{d^4k_1}{\pi^2i} \hspace*{-0.1cm}
\int\hspace*{-0.1cm}\frac{d^4k_2}{\pi^2i} \hspace*{0.1cm}
\tilde F_B\biggl\{-6\biggl[k_1^2+k_2^2+(k_1+k_2)^2\biggr]\biggr\}
\\
&\times&
\tilde F_B\biggl\{-6\biggl[(k_1+q)^2+(k_2-q)^2+(k_1+k_2)^2\biggr]\biggr\}
\nonumber\\
&\times&
\frac{\tilde F_\pi\biggl\{-(k_2-q/2)^2\biggr\}}
{[-k_1v-\bar\Lambda_{q_1q_2}]}
\cdot\frac{1}{4}{\rm tr}\biggl[\Gamma_2^i S_{q_2}(k_1+k_2)\Gamma_2^f
S_{q_1}(k_2-q)\gamma^5 S_{q_1}(k_2)\biggr]
\nonumber
\end{eqnarray}
where $\lambda_\pi$, $\lambda_{B^i}$ and $\lambda_{B^f}$ are the
flavor matrices of the pion, the initial and the final baryons, respectively.

Here $S_q(k)=1/(m_q - \not\!\!k)$ is the light quark propagator $(q=u,d,s)$.
The parameter $\bar\Lambda_{q_1q_2}=M_{Qq_1q_2}-m_Q$ denotes the difference
between the heavy baryon mass $M_{Qq_1q_2}$ and the heavy quark mass $m_Q$.
All dimensional parameters are expressed in units of $\Lambda_B$.
The integrals are calculated in the Euclidean region both
for internal and external momenta. Finally, the results for the physical
region are obtained by analytic continuation of the external momenta after
the internal momenta have been integrated out.

As an illustration of our calculational procedure we first evaluate the
matrix element
Eq. (\ref{int}) in the simplified case where the pion has a local coupling to
its constituent quarks, i.e. where the vertex $\pi q \bar q$ form factor
$\tilde F_\pi(k^2)\equiv 1$. As it turns out this is already a good
appriximation. For example,
we have calculated the integral Eq. (\ref{int}) with the baryonic vertex
function being chosen in the Gaussian form for two cases:
(1) $\tilde F_\pi \equiv $1 and (2) $\tilde F_\pi = \exp(-k^2/\Lambda_\pi^2)$,
$\Lambda_\pi=$1 GeV. The results differ from each other by $O(10\%)$.

In the calculation of (\ref{int}) with $\tilde F_\pi(k^2)\equiv 1$
we use the $\alpha$-parametrization for quark propagators
and the Laplace transform for the vertex function
\begin{eqnarray}
\frac{1}{A}=\int\limits_0^\infty\hspace*{-0.1cm}d\alpha e^{-\alpha A},
\hspace{1cm}
\tilde F_B(6X)=\int\limits_0^\infty\hspace*{-0.1cm}ds \tilde F^L_B(6s)e^{-sX}
\end{eqnarray}
\begin{eqnarray}\label{calc}
I^{if}(v,q)&=&
\int\limits_0^\infty \hspace*{-0.1cm}ds_1 \tilde F_B^L(6s_1)
\int\limits_0^\infty \hspace*{-0.1cm}ds_2 \tilde F_B^L(6s_2)e^{2s_2q^2}
\int\limits_0^\infty \hspace*{-0.1cm}d^4 \alpha
e^{\alpha_3\bar\Lambda-(\alpha_1+\alpha_4)m_{q_1}^2-\alpha_2m_{q_2}^2}
\nonumber\\
&\times&\frac{1}{4}{\rm tr}\biggl[\Gamma_2^i
\biggl(m_{q_2}-\frac{\not\!\partial_1+\not\!\partial_2}{2}\biggr)\Gamma_2^f
\biggl(m_{q_1}-\frac{\not\!\partial_2}{2}-\not\! q\biggr)\gamma^5
\biggl(m_{q_1}-\frac{\not\!\partial_2}{2}\biggr)\biggr]
\int\hspace*{-0.1cm}\frac{d^4k_1}{\pi^2i} \hspace*{-0.1cm}
\int\hspace*{-0.1cm}\frac{d^4k_2}{\pi^2i} \hspace*{0.1cm}
e^{kAk-2kB}
\nonumber \\
&=&\int\limits_0^\infty \hspace*{-0.1cm}ds_1 \tilde F_B^L(6s_1)
\int\limits_0^\infty \hspace*{-0.1cm}ds_2 \tilde F_B^L(6s_2)e^{2s_2q^2}
\int\limits_0^\infty \hspace*{-0.1cm}d^3\alpha
e^{\alpha_3\bar\Lambda-(\alpha_1+\alpha_4)m_{q_1}^2-\alpha_2m_{q_2}^2}
\nonumber\\
&\times&\frac{1}{4}{\rm tr}\biggl[\Gamma_2^i
\biggl(m_{q_2}-\frac{\not\!\partial_1+\not\!\partial_2}{2}\biggr)\Gamma_2^f
\biggl(m_{q_1}-\frac{\not\!\partial_2}{2}-\not\! q\biggr)\gamma^5
\biggl(m_{q_1}-\frac{\not\!\partial_2}{2}\biggr)\biggr]
\frac{e^{-BA^{-1}B}}{|A|^2}
\nonumber
\end{eqnarray}
Here
\begin{eqnarray}
k A k - 2 k B = \sum\limits_{i,j=1}^{2} k_i A_{ij} k_j -
2\sum\limits_{i=1}^{2} k_i B_i, \hspace*{1cm}
\not\!\partial_i = \frac{\partial}{\partial\not\!\!B_i}
\end{eqnarray}
\[A_{ij}=\left(
\begin{array}{ll}
 \mbox{$2(s_1+s_2)+\alpha_2$}  &  \hspace*{.5cm}  \mbox{$s_1+s_2+\alpha_2$}   \\
     &\\
 \mbox{$s_1+s_2+\alpha_2$}     &  \hspace*{.5cm}
\mbox{$2(s_1+s_2)+\alpha_1+\alpha_2+\alpha_4$}
\end{array}
\right) \]
$$
B_1=-s_2q-\alpha_3v/2 \hspace{1cm}  B_2=(s_2+\alpha_1)q
$$
The kinematics of the one-pion transitions allows one to make use of the
approximations:
\begin{eqnarray}
q^2=m_\pi^2\approx 0,
\hspace*{1cm}
qv=\frac{1}{2m_i}(m_i^2-m_f^2+m^2_\pi)\approx 0
\end{eqnarray}
where $m_\pi$, $m_i$ and $m_f$ are the masses of the pion, the
initial and the final baryons, respectively, divided by $\Lambda_B$.
Then, by making the variable replacement $\alpha_i\to(s_1+s_2)\alpha_i$
and by using
$\Gamma_2^i = \gamma_\mu$ and $\Gamma_2^f=\gamma_5$ the overlap integral
can be seen to be proportional to $q^\mu$ such that
\begin{eqnarray}
I^\mu(v,q)=q^\mu J
\end{eqnarray}
with
$$
J=\int\limits_0^\infty\hspace*{-0.1cm}\frac{d^3\alpha\alpha_1}{|A|^2}
\tilde F_B^2(6z)\biggl\{m_{q_1}m_{q_2}+
\alpha_3\frac{\partial z}{\partial \alpha_3} [A^{-1}_{12}+A^{-1}_{22}]
\biggl[1 + \frac{(1+\alpha_1)A^{-1}_{22}-A^{-1}_{12}}{2}\biggr]
-\frac{\alpha_3^2}{4}A^{-1}_{12}[A_{11}^{-1}+A_{12}^{-1}]\biggr\}
$$
$$
z=\frac{\alpha_3^2}{4}A^{-1}_{11}-\alpha_3\bar\Lambda
+\alpha_1m^2_{q_1}+\alpha_2m^2_{q_2}
$$
\[A_{ij}=\left(
\begin{array}{ll}
 \mbox{$2+\alpha_2$}      &    \hspace*{.2cm}    \mbox{$ 1+\alpha_2$}   \\
 &\\
 \mbox{$ 1+\alpha_2$}     &    \hspace*{.2cm}    \mbox{$2+\alpha_1+\alpha_2$}
\end{array}
\right),
\hspace*{.5cm}
A^{-1}_{ij}=\frac{1}{|A|}\left(
\begin{array}{ll}
 \mbox{$2+\alpha_1+\alpha_2$} &    \hspace*{.2cm}    \mbox{$-(1+\alpha_2)$}   \\
 &\\
 \mbox{$-(1+\alpha_2)$}       &    \hspace*{.2cm}    \mbox{$2+\alpha_2$}
\end{array}
\right) \]

The last integral may be evaluated numericaly for any given function
$\tilde F_B$. Here we will use a Gaussian vertex functions
both for baryons and the pion in Eq.~(\ref{int}).

In order to make contact with experimental numbers let us first
define a set of coupling constants describing the one-
pion transitions. For the transitions discussed in this paper the
coupling constants are defined through the expansion of the
the invariant one-pion transition matrix elements. One has \cite{Tawfiq}
\begin{eqnarray}\label{matrix_el}
M_{inv}^\pi(\Sigma_c\to\Lambda_c\pi) &=&
\frac{1}{\sqrt{3}}g_{\Sigma_c\Lambda_c\pi}
I_1  \bar u(v^\prime) \not\! q_{\perp}\gamma_5 u(v)
\hspace*{.5cm}
\underline{\mbox{p-wave transition}}
\nonumber \\
M_{inv}^\pi(\Sigma_c^*\to\Lambda_c\pi) &=&
g_{\Sigma^{*}_c\Lambda_c\pi} I_1
\bar{u}(v^\prime)q_{\perp\mu}u^{\mu}(v)
\hspace*{.5cm}
\underline{\mbox{p-wave transition}}
\nonumber \\
& &\\
M_{inv}^\pi(\Lambda_{c1}\to\Sigma_c\pi) &=&  f_{\Lambda_{c1}\Sigma_c\pi} I_3
\bar{u}(v^{\prime})u(v)
\hspace*{.5cm}
\underline{\mbox{s-wave transition}}
\nonumber \\
M_{inv}^\pi(\Lambda_{c1}^*\to\Sigma_c\pi) &=& \frac{1}{\sqrt{3}}
f_{\Lambda_{c1}^{*}\Sigma_c\pi} I_3
\bar{u}(v^{\prime})\gamma_5\not\! q_{\perp}u^{\mu}(v)q_{\perp \mu}
\hspace*{.5cm}
\underline{\mbox{d-wave transition}}
\nonumber
\end{eqnarray}
where the $I_1$ and $I_3$ are the flavor factors
which are directly connected with flavor coefficients
$C_{\rm flavor}$ (see Eq. (\ref{vertex})) via relations
$I_i = f_i \cdot C_{\rm flavor}$, $i=1$ or $3$. The sets of $I_i$ and $f_i$
are given in TABLE II.
We have also indicated the orbital angular momentum of the pion in
Eq. (\ref{matrix_el}). The transversity in Eq. (\ref{matrix_el})
is defined with regard of the velocity $v=p/m$ of the decaying baryon,
i.e. $q^{\perp}_\mu = q_\mu - v_\mu (q\cdot v)$. In fact, to leading order
in the HQET expansion one has $v=v^\prime$. In general, however,
$v\not = v^\prime$ from momentum conservation. By keeping track of
momentum conservation in Eq. (\ref{matrix_el}) using $v\not = v^\prime$
one is including a part of the nonleading effects. The structure of
the covariants in Eq. (\ref{matrix_el}) is patterned after the leading order
HQET result which predicts
$g_{\Sigma_c\Lambda_c\pi}=g_{\Sigma^{*}_c\Lambda_c\pi}=g$ \cite{Yan}.
It is an easy exercise to rewrite e.g. the p-wave
$\not\! q_{\perp}\gamma_5$-coupling in terms of the usual
$\gamma_5$-coupling.
The expression for $g$ is written as
\begin{eqnarray}
g&=&\frac{1}{\Lambda_B}\cdot\frac{g_\pi}{\sqrt{2}}\cdot
\frac{R_{\Sigma\Lambda\pi}}{\sqrt{R_\Lambda}\sqrt{R_\Sigma}}\\
R_{\Sigma\Lambda\pi}&=&
\int\limits_0^\infty
\hspace*{-.1cm}
\frac{d\alpha\alpha^2}{1+\alpha+t}
\int\limits_0^\infty \hspace*{-.1cm}d\beta
\int\limits_0^1 \hspace*{-.1cm}
\frac{d\theta\theta\exp(\Delta_1) }{\frac{3}{4}+\alpha+
\alpha^2\theta(1-\theta)+t(1+\alpha(1-\theta))}
\nonumber\\
&\times&\biggl\{m_{q_1}m_{q_2}+\beta^2
\frac{\frac{1}{4}+\frac{\alpha}{2}+\alpha^2\theta(1-\theta)}{(1+\alpha+t)^2}
+\frac{1}{32[\frac{3}{4}+\alpha+\alpha^2\theta(1-\theta)+t(1+\alpha(1-\theta))]}
\biggr\}\nonumber\\
R_{\Lambda}&=&
\int\limits_0^\infty \hspace*{-.1cm}
\frac{d\alpha\alpha}{(1+\alpha)^2}
\int\limits_0^\infty \hspace*{-.1cm} d\beta\beta
\int\limits_0^1 \hspace*{-.1cm} d\theta \exp(\Delta_2)
\biggl\{m_{q_1}m_{q_2}+\beta^2
\frac{\frac{5}{4}+\frac{3}{2}\alpha+\alpha^2\theta(1-\theta)}{(1+\alpha)^2}
-\frac{\bar\Lambda\beta}{1+\alpha}\biggr\}\nonumber\\
R_{\Sigma}&=& \int\limits_0^\infty \hspace*{-.1cm}
\frac{d\alpha\alpha}{(1+\alpha)^2}
\int\limits_0^\infty d\beta\beta
\int\limits_0^1 d\theta \exp(\Delta_2)
\biggl\{m_{q_1}m_{q_2}+\beta^2
\frac{\frac{3}{4}+\alpha+\alpha^2\theta(1-\theta)}{(1+\alpha)^2}
-\frac{\bar\Lambda\beta}{2(1+\alpha)}\biggr\}\nonumber\\
&\Delta_1&=-24\biggl\{
\alpha[m_{q_1}^2\theta+m_{q_2}^2(1-\theta)]
+\beta(\beta-2\bar\Lambda)\frac{\frac{3}{4}+\alpha+\alpha^2\theta(1-\theta)+
t(\frac{1}{2}+\alpha(1-\theta))}{1+\alpha+t}\biggr\}
\nonumber\\
&\Delta_2&\equiv\Delta_1|_{t=0},
\hspace*{1cm}t=(\Lambda_B/\Lambda_\pi\sqrt{24})^2
\nonumber
\end{eqnarray}

One can then go on and calculate the one-pion decay rates using the general
formula
\begin{eqnarray}\label{rate}
\Gamma = \frac{1}{2J+1} \quad \frac{ \mid \vec{q} \mid}{8 \pi
M_{B_Q}^{2}}\sum_{spins} \mid M^{\pi}_{inv} \mid^{2}
\end{eqnarray}
where $\mid \vec{q} \mid $ is the pion momentum in the rest frame
of the decaying baryon. In terms of the above coupling constants one
obtains
\begin{eqnarray}
\Gamma\left( \Sigma_c \rightarrow \Lambda_c \pi \right)
&=& g^2I_1^2
\frac{{\mid \vec{q} \mid}^3}{6\pi} \frac{M_{\Lambda_c}}{M_{\Sigma_c}}
\label{prate}\\
\Gamma\left( \Sigma^*_c \rightarrow \Lambda_c \pi \right)
&=& g^2I_1^2
\frac{{\mid \vec{q} \mid}^3}{6\pi} \frac{M_{\Lambda_c}}{M_{\Sigma^*_c}}
\label{p(*)rate}\\
\Gamma\left(\Lambda_{c1}  \rightarrow  \Sigma_c \pi  \right)
     &=& f^2_{\Lambda_{c1}\Sigma\pi} I_3^2
         \frac{\mid \vec{q} \mid}{2\pi} \frac{M_{\Sigma_c}}{M_{\Lambda_{c1} }}
\label{srate}\\
\Gamma\left(\Lambda^{*}_{c1}  \rightarrow  \Sigma_c \pi \right)
   &=& f^2_{\Lambda^{*}_{c1}\Sigma\pi} I_3^2
    \frac{{\mid \vec{q} \mid}^5}{18\pi} \frac{M_{\Sigma_c}}
    {M_{\Lambda^{*}_{c1} }}
\label{drate}
\end{eqnarray}

We use different values for the parameter $\bar\Lambda_{q_1q_2}$
for baryons containing only nonstrange light quarks and one or two strange
quarks: $\bar\Lambda$, $\bar\Lambda_s$ and $\bar\Lambda_{ss}$, respectively.
For the time being we shall avoid the appearance of unphysical imaginary
parts in the Feynman diagrams by imposing the following condition:
the baryon mass must be less than the sum of constituent quark masses.
In the case of heavy-light baryons this restriction implies that the
parameter $\bar\Lambda_{q_1q_2}$ must be less than the sum of light
quark masses. The last constraint serves as the upper limit for our choices
of the parameter $\bar\Lambda_{q_1q_2}$.

Let us now specify our model parameters. The coupling constants $g_{B_Q}$
and $g_\pi$ in Eqs. (1) and (2) are calculated
from {\it the compositeness condition} (see, ref.~\cite{PRD97}), which means
that the renormalization constant of the hadron wave function is set equal to
zero $Z_H=1-g_H^2\Sigma^\prime_H(M_H)=0$ where $\Sigma_H$ is the hadron
(charm baryon and pion) mass operator. We thus remain with the cutoff
parameters $(\Lambda_B, \Lambda_\pi)$ and parameters $(\bar\Lambda$,
$\bar\Lambda_s, \bar\Lambda_{ss})$ as the adjustable parameters
in our approach. The masses of the $u$ and the $d$ quarks are set equal
($m_u=m_d=m_q$). The value of $m_q$ is determined from an analysis of
nucleon data: $m_q$=420 MeV. The pion cutoff parameter $\Lambda_\pi=$ 1 GeV
is fixed from the description of low-energy pion observables (constants
$f_\pi$ and $g_{\pi\gamma\gamma}$, electromagnetic radii and form factors
defining the transitions $\pi\to\pi\gamma$ and $\pi\to\gamma\gamma^*$)
\cite{YadFiz}-\cite{PIGG}.
The parameters $\Lambda_{B_Q}$, $m_s$, $\bar\Lambda$ are taken from
the analysis of the $\Lambda^+_c\to\Lambda^0+e^+ +\nu_e$ decay data.
To reproduce the present average value of
$B(\Lambda_c^+\to\Lambda e^+ \nu_e$) = 2.2 $\%$ we
use the following values for our parameters:
$\Lambda_{B_Q}$=1.8 GeV, $m_s$=570 MeV and $\bar\Lambda$=600 MeV.
The values of the unknown parameters $\bar\Lambda_s$ and $\bar\Lambda_{ss}$
are determined from the relations $\bar\Lambda_s = \bar\Lambda + (m_s - m)$
and $\bar\Lambda_{ss} = \bar\Lambda + 2(m_s - m)$, which give
$\bar\Lambda_s$ = 750 MeV and $\bar\Lambda_{ss}$ = 900 MeV. Using the values
of $\Lambda_{B_Q}$=1.8 GeV and $\bar\Lambda$=600 MeV one can describe
the decay $\Lambda_b^0\to\Lambda_c^+ e^- \bar\nu_e$ decay: the width
$\Gamma(\Lambda_b^0\to\Lambda_c^+e^-\bar\nu_e)=5.06\times 10^{10}s^{-1}$
and the slope of the $\Lambda_b$ Isgur-Wise function $\rho^2 = 1.44$.
Finally, the mass values of the charm baryon states including current
experimental uncertainties, are taken from TABLE I \cite{PDG,Chiladze}.
For $\Xi_c^{+\prime}$ and $\Xi_c^{0\prime}$ we use mass value with
theoretical uncertainty $m_{\Xi_c^{\prime}} = 2600 \pm 30$ MeV.
For the pion masses we take their experimental values
$m_{\pi^\pm} = $ 139.6 MeV and $m_{\pi^0}$ = 135 MeV~\cite{PDG}.

We now present our numerical results for the strong charm
baryon-pion couplings and for the one-pion decay rates.
In TABLE III we list our results for one-pion coupling constants.
For comparison we give also the results of Light-Front (LF) Quark Model
\cite{Tawfiq} which have been only available up to now.
One can see that the values of the $P$-wave coupling constants
are in qualitive
agreement with the Light-Front quark model prediction. However, we disagree
on the values of the $S$-wave and $D$-wave coupling constants
$f_{\Lambda_{c1}\Sigma_c\pi}$ and $f_{\Lambda_{c1}^{*}\Sigma_c\pi}$.
The disagreement can be traced to different choices of the momentum
distribution of the light quarks in the charm baryon. The smaller values of
$f_{\Lambda_{c1}\Sigma_c\pi}$ and  $f_{\Lambda_{c1}^{*}\Sigma_c\pi}$
are welcome from comparison the results for exclusive one-pion rates of the
$\Lambda_{c1}$ and $\Lambda_{c1}^{*}$ (see, TABLE IV) to the experimental
data \cite{PDG}. It is seen that our predictions are consistent with current
experimental estimations whereas the Light-Front model results lay above
the experimental rates. Also our predictions for $S$-wave and $D$-wave
transitions are preferable if one sums up the three exclusive one-pion rates
of the $\Lambda_{c1}$ and $\Lambda_{c1}^{*}$ and compares the sums to the
total experimental rates $\Gamma(\Lambda_{c1})=3.6^{+2.0}_{-1.3}$ MeV and
$\Gamma({\Lambda^{*}_{c1}}) <1.9$ MeV. From the results of our model
calculation we obtain $\Gamma(\Lambda_{c1}) > 2.6 \pm 0.3 $ MeV and
$\Gamma(\Lambda_{c1}^{*}) > 0.25 \pm 0.03 $ MeV consistent with the
experimental total rate of $\Lambda_{c1}$ and lower limit for rate of
$\Lambda_{c1}^{*}$ whereas the Light-Front model has
$\Gamma(\Lambda_{c1})>6.49$ MeV and $\Gamma(\Lambda_{c1}^{*})>2.19 $ MeV
which lay above the experimental rates. One can hope that more precise
experimental study of strong decays of excited $\Lambda_{c1}$ baryons in
the near future can test the predictions of our approach and the LF quark
model. All our results for the one-pion decay rates of charm baryons are
collected in TABLE IV. The uncertainties for the calculated rates reflect
the experimental errors in the charm baryon masses (see, TABLE I). For
comparison we have also listed the predictions of the Light-Front quark
model \cite{Tawfiq} and experimental results, where available. One can only
hope that there will be more precise data on the one-pion transitions of the
excited $\Lambda_{c1}$ baryon states to the ground states in the near future
such that one can perform a more detailed comparison with the model
predictions of dynamical models such as described in our approach and
in the LF quark model.

Let us add a comment of the relation of our approach to the chiral
invariant coupling method used for example in \cite{Yan}.
The chiral formalism implies that all one-pion coupling factors are
proportional to the factor $1/f_\pi$ associated with the pion field.
In our approach the corresponding factor emerges in the following way.
The pion-quark-antiquark coupling $g_\pi$ can be seen to obey the
Goldberger-Treiman relation $g_\pi \cdot f_\pi = 2m_q$ with an accuracy of
a few percent. Hence our approach agrees with the chiral approach \cite{Yan}
in that the pion leptonic constant $f_\pi$ effectively appears as
a dimensional parameter in the coupling factors.

In conclusion, we have calculated strong one-pion decays of charm
baryons. We have obtained predictions for the values of couplings of charm
baryons with pions and for the rates of the two-body transitions
$B^i_c(p)\to B^f_c(p^\prime)+\pi(q)$. We have compared our results with
data obtained with the use Light-Front Quark Model \cite{Tawfiq}.
As a next step we plan to study one-photon transitions
between charm baryons. Also we intend to extend our results
to the bottom baryon sector.

\vspace*{1cm}
\noindent
{\bf Acknowledgments}

\vspace*{.5cm}
\noindent
M.A.I, V.E.L and A.G.R thank Mainz University for the hospitality
where a part of this work was completed. This work was supported
in part by the Heisenberg-Landau Program, by the Russian Fund of
Fundamental Research (RFFR) under contract 96-02-17435-a and by
the BMBF (Germany) under contract 06MZ566. J.G.K. aknowledges
partial support by the BMBF (Germany) under contract 06MZ566.
V.E.L. thanks the Russian Federal Program "Integration of
Education and Fundamental Science" for partial support.

\vspace*{2cm}
\centerline{\bf List of Tables}
\noindent
{\bf TABLE I}
Quantum numbers of charm baryons
($\lambda_u$ = diag\{1,0,0\}, $\lambda_d$ = diag\{0,1,0\})

\vspace*{.2cm}
\noindent
{\bf TABLE II}
Flavor coefficients $I_1, I_3$ and $f_1, f_3$.

\vspace*{.2cm}
\noindent
{\bf TABLE III}
Charm baryon-pion couplings.

\vspace*{.2cm}
\noindent
{\bf TABLE IV}
Decay rates $\Gamma$ (in MeV) for charm baryon states.

\vspace*{1cm}
\centerline{\bf List of Figures}

\noindent
{\bf FIG. I}
One-pion charm baryon transition: $\Sigma_c\to\Lambda_c\pi$ decay.

\newpage
\begin{center}
{\bf TABLE I}
\end{center}
\def\arraystretch{1.25}
\begin{center}
\begin{tabular}{|c|c|c|c|c|c|}
\hline
Baryon&$\;J^P\;$&Quark Content& $\Gamma_1\otimes C\Gamma_2$
& $\; \lambda_{B_Q}\;$& $\;$Mass (MeV) \cite{PDG,Chiladze} $\;$\\
\hline
$\Lambda_c^+$    & ${\frac{1}{2}}^+$&c[ud] &$I\otimes C\gamma^5$ & $i\lambda_2/2$ & $2284.9\pm 0.6$\\
\hline
$\Xi_c^+$        & ${\frac{1}{2}}^+$&c[us] &$I\otimes C\gamma^5$ & $i\lambda_5/2$ & $2465.6\pm 1.4$\\
\hline
$\Xi_c^0$        & ${\frac{1}{2}}^+$&c[ds] &$I\otimes C\gamma^5$ & $i\lambda_7/2$ & $2470.3\pm 1.8$\\
\hline
$\Xi_c^{+\prime}$& ${\frac{1}{2}}^+$&c\{us\} &$\gamma^\mu\gamma^5 \otimes C\gamma_\mu$ & $\lambda_4/(2\sqrt{3})$ & $2600\pm 30$\\
\hline
$\Xi_c^{0\prime}$& ${\frac{1}{2}}^+$&c\{ds\} &$\gamma^\mu\gamma^5 \otimes C\gamma_\mu$ & $\lambda_6/(2\sqrt{3})$ & $2600\pm 30$\\
\hline
$\Sigma_c^{++}$  & ${\frac{1}{2}}^+$&c\{uu\} &$\gamma^\mu\gamma^5 \otimes C\gamma_\mu$ & $\lambda_u/\sqrt{6}$ & $2452.8\pm 0.6$\\
\hline
$\Sigma_c^+$     & ${\frac{1}{2}}^+$&c\{ud\} &$\gamma^\mu\gamma^5 \otimes C\gamma_\mu$ & $\lambda_1/(2\sqrt{3})$ & $2453.6\pm 0.9$\\
\hline
$\Sigma_c^0$     & ${\frac{1}{2}}^+$&c\{dd\} &$\gamma^\mu\gamma^5 \otimes C\gamma_\mu$ & $\lambda_d/\sqrt{6}$ & $2452.2\pm 0.6$\\
\hline
$\Xi_c^{*+}$     & ${\frac{3}{2}}^+$&c\{us\} &$I \otimes C\gamma_\mu$ & $\lambda_4/2$ & $2644.6\pm 2.1$\\
\hline
$\Xi_c^{*0}$     & ${\frac{3}{2}}^+$&c\{ds\} &$I \otimes C\gamma_\mu$ & $\lambda_6/2$ & $2643.8\pm 1.8$\\
\hline
$\Sigma_c^{*++}$ & ${\frac{3}{2}}^+$&c\{uu\} &$I \otimes C\gamma_\mu$ & $\lambda_u/\sqrt{2}$ & $2519.4\pm 1.5$\\
\hline
$\Sigma_c^{*0}$    & ${\frac{3}{2}}^+$&c\{dd\} &$I \otimes C\gamma_\mu$ & $\lambda_d/\sqrt{2}$ & $2517.5\pm 1.4$\\
\hline
$\Lambda_{c1}$   & ${\frac{1}{2}}^-$  &c[ud]  &$\not\!\partial_{\xi_1}\gamma_5 \otimes C\gamma^5$ & $i\lambda_2/(2\sqrt{3})$
& $2593.9\pm 0.8$\\
\hline
$\Lambda_{c1}^*$ & ${\frac{3}{2}}^-$  &c[ud]  &$\partial^\mu_{\xi_1} \otimes C\gamma^5$ & $i\lambda_2/2$ & $2626.6\pm 0.8$\\
\hline
$\Xi_{c1}^*$     & ${\frac{3}{2}}^-$  &c[us]  &$\partial^\mu_{\xi_1} \otimes C\gamma^5$ & $i\lambda_5/2$ & 2815\\
\hline
\end{tabular}
\end{center}

\newpage
\begin{center}
{\bf TABLE II}
\end{center}
 \renewcommand{\baselinestretch}{1.2}
 \small \normalsize
 \begin{center}
 \begin{tabular}{|c|c|c||c|c|c|}
 \hline \hline
 Decay mode & $I_1$  &  $f_1$ & Decay mode & $I_3$ & $f_3$ \\
 \hline\hline
$\Sigma^{+}_{c} \rightarrow \Lambda_c\pi^{0} $  & $1$ & $\sqrt{3}/2$ &   $\Lambda_{c1}(2593) \rightarrow \Sigma^{0}_c\pi^{+} $ & $1$ & $3/2$ \\
\hline
$\Sigma^{0}_{c} \rightarrow \Lambda_c\pi^{-} $  & $1$ & $\sqrt{3}/2$ &   $\Lambda_{c1}(2593) \rightarrow \Sigma^{+}_c\pi^{0} $ & $1$ & $3/2$ \\
\hline
$\Sigma^{++}_{c} \rightarrow \Lambda_c\pi^{+}$  & $1$ & $\sqrt{3}/2$ &   $\Lambda_{c1}(2593) \rightarrow \Sigma^{++}_c\pi^{-}$ & $1$ & $3/2$ \\
\hline
$\Sigma^{*0}_{c} \rightarrow \Lambda_c\pi^{-}$  & $1$ & $1/2$        &   $\Xi^{*}_{c1}(2815) \rightarrow \Xi^{*0}_c\pi^{+}   $ &  $1/\sqrt{2}$ & $1/2$ \\
\hline
$\Sigma^{* ++}_{c} \rightarrow \Lambda_c\pi^{+}$& $1$ & $1/2$        &   $\Xi^{*}_{c1}(2815) \rightarrow \Xi^{*+}_c\pi^{0}   $ & $1/2$ & $1/2$ \\
\hline
$\Xi^{*0}_{c} \rightarrow \Xi^{0}_c\pi^{0} $    & $1/2$ & $1/2$        &   $\Lambda^{*}_{c1}(2625) \rightarrow \Sigma^{0}_c\pi^{+} $ & $1$ & $\sqrt{3}/2$  \\
\hline
$\Xi^{*0}_{c} \rightarrow \Xi^{+}_c\pi^{-} $    & $1/\sqrt{2}$ &  $1/2$ & $\Lambda^{*}_{c1}(2625) \rightarrow \Sigma^{+}_c\pi^{0} $ & $1$ & $\sqrt{3}/2$ \\
\hline
$\Xi^{*+}_{c} \rightarrow \Xi^{0}_c\pi^{+} $    & $1/\sqrt{2}$ &  $1/2$ & $\Lambda^{*}_{c1}(2625) \rightarrow \Sigma^{++}_c\pi^{-}$ & $1$ & $\sqrt{3}/2$ \\
\hline
$\Xi^{*+}_{c} \rightarrow \Xi^{+}_c\pi^{0} $    & 1/2        &    $1/2$ & $\Xi^{*}_{c1}(2815) \rightarrow \Xi^{0\prime}_c\pi^{+}  $ & $1/\sqrt{2}$ & $\sqrt{3}/2$ \\
\hline
                                                &            &          & $\Xi^{*}_{c1}(2815) \rightarrow \Xi^{+\prime}_c\pi^{0}  $ & $1/2$ & $\sqrt{3}/2$\\
\hline \hline
\end{tabular}
\end{center}

\vspace*{1.5cm}
\begin{center}
{\bf TABLE III}
\end{center}
\def\arraystretch{1.25}
\begin{center}
\begin{tabular}{|c|c|c|}
\hline\hline
Coupling & Our & Ref. \cite{Tawfiq}\\
\hline\hline
$g_{\Sigma_c\Lambda_c\pi}$ & 8.88 GeV$^{-1}$ & 6.81 GeV$^{-1}$   \\
$g_{\Xi_c^*\Xi_c\pi}$ & 8.34 GeV$^{-1}$ &                        \\
\hline
$f_{\Lambda_{c1}\Sigma_c\pi}$ & 0.52 & 1.16                      \\
$f_{\Xi_{c1}^*\Xi_c^*\pi}$ & 0.32 &                              \\
\hline
$f_{\Lambda_{c1}^*\Sigma_c\pi}$ & 21.5 GeV$^{-2}$ &96.0 GeV$^{-2}$\\
$f_{\Xi_{c1}^*\Xi_c^\prime\pi}$ & 20 GeV$^{-2}$ &  \\
\hline
\end{tabular}
\end{center}

\newpage
\begin{center}
{\bf TABLE IV}
\end{center}
 \renewcommand{\baselinestretch}{1.2}
 \small \normalsize
 \begin{center}
 \begin{tabular}{|c|c|c|c|}
 \hline \hline
 $B_Q\rightarrow B^{\prime}_{Q}\pi$ & Our & Ref. \cite{Tawfiq} &
 Experiment  \\
 \hline\hline
\multicolumn{4}{|l|}{P-wave transitions} \\
\hline
$\Sigma^{+}_{c} \rightarrow \Lambda_c\pi^{0} $ & $ 3.63\pm 0.27 $ & $1.70$ & $  $\\
$\Sigma^{0}_{c} \rightarrow \Lambda_c\pi^{-} $ & $ 2.65\pm 0.19 $ & $1.57$ & $  $\\
$\Sigma^{++}_{c} \rightarrow \Lambda_c\pi^{+}$ & $ 2.85\pm 0.19 $ & $1.64$ & $  $\\
\hline
$\Sigma^{*0}_{c} \rightarrow \Lambda_c\pi^{-}$ & $ 21.21 \pm 0.81$ & $ 12.40$ &
$ 13.0^{+3.7}_{-3.0} $   \\
$\Sigma^{* ++}_{c} \rightarrow \Lambda_c\pi^{+}$ &  $ 21.99\pm 0.87 $ & $ 12.84$ &
$ 17.9^{+3.8}_{-3.2} $   \\
\hline
$\Xi^{*0}_{c} \rightarrow \Xi^{0}_c\pi^{0} $ &  $ 1.01 \pm 0.15  $ & $ 0.72$
& \\
$\Xi^{*0}_{c} \rightarrow \Xi^{+}_c\pi^{-} $ &  $ 2.11 \pm 0.29$ & $ 1.16$ &
$ \Gamma(\Xi^{*0})< 5.5 $   \\
\hline
$\Xi^{*+}_{c} \rightarrow \Xi^{0}_c\pi^{+} $ &  $ 1.78 \pm 0.33  $ & $ 1.12$ & \\
$\Xi^{*+}_{c} \rightarrow \Xi^{+}_c\pi^{0} $ &  $ 1.26 \pm 0.17  $ & $ 0.69$ &
$ \Gamma(\Xi^{*+})< 3.1 $ \\
\hline \hline
\multicolumn{4}{|l|}{S-wave transitions} \\
\hline
$\Lambda_{c1}(2593) \rightarrow \Sigma^{0}_c\pi^{+} $ & $0.83\pm 0.09  $ & $ 2.61$
& $0.86^{+0.73}_{-0.56}$ \\
$\Lambda_{c1}(2593) \rightarrow \Sigma^{+}_c\pi^{0} $ & $0.98\pm 0.12  $ & $ 1.73$
& $ \Gamma({\Lambda_{c1}}) = 3.6^{+2.0}_{-1.3}$  \\
$\Lambda_{c1}(2593) \rightarrow \Sigma^{++}_c\pi^{-} $ & $0.79\pm 0.09$ & $ 2.15$
& $0.86^{+0.73}_{-0.56}$ \\
\hline
$\Xi^{*}_{c1}(2815) \rightarrow \Xi^{*0}_c\pi^{+} $ &  $ 0.91\pm 0.03  $ & $ 4.84$
& \\
$\Xi^{*}_{c1}(2815) \rightarrow \Xi^{*+}_c\pi^{0} $ &  $ 0.48\pm 0.02  $ & $ 2.38$
& $ \Gamma({\Xi^{*}_{c1}})<2.4 $ \\
\hline \hline
\multicolumn{4}{|l|}{D-wave transitions} \\
\hline
$\Lambda^{*}_{c1}(2625) \rightarrow \Sigma^{0}_c\pi^{+} $ & $ 0.080\pm 0.009 $
& $ 0.77$ & $< 0.13$\\
$\Lambda^{*}_{c1}(2625) \rightarrow \Sigma^{+}_c\pi^{0} $ & $ 0.095\pm 0.012 $
& $ 0.69$ & $\Gamma({\Lambda^{*}_{c1}}) <1.9 $  \\
$\Lambda^{*}_{c1}(2625) \rightarrow \Sigma^{++}_c\pi^{-} $& $ 0.076\pm 0.009 $
& $ 0.73$ & $<0.15$\\
\hline
$\Xi^{*}_{c1}(2815) \rightarrow \Xi^{0\prime}_c\pi^{+} $ & $0.46\pm 0.39$
& $ 0.30$ &$ $  \\
$\Xi^{*}_{c1}(2815) \rightarrow \Xi^{+\prime}_c\pi^{0} $ & $0.25\pm 0.21$
& $ 0.15$ & $ \Gamma({\Xi_{c1}^*}) < 2.4$\\
\hline \hline
\end{tabular}
 \renewcommand{\baselinestretch}{1}
 \small \normalsize
 \end{center}

\newpage
\unitlength=1.00mm
\special{em:linewidth 0.4pt}
\linethickness{0.4pt}
\begin{picture}(115.00,124.00)
\put(50.00,80.00){\circle*{5.20}}
\put(95.00,80.00){\circle*{5.20}}
\put(73.00,103.00){\circle*{4.00}}
\put(75,128){\makebox(0,0)[cc]{\LARGE$\bf \pi^+$}}
\put(73.00,64.00){\makebox(0,0)[cc]{\Large$\bf c$}}
\put(30.00,86.00){\makebox(0,0)[cc]{\Large$\bf \Sigma_c^{++}$}}
\put(115.00,86.00){\makebox(0,0)[cc]{\Large$\bf \Lambda_c^+$}}
\put(73.00,84.00){\makebox(0,0)[cc]{\Large$\bf u$}}
\put(59.00,94.00){\makebox(0,0)[cc]{\Large$\bf u$}}
\put(87.00,94.00){\makebox(0,0)[cc]{\Large$\bf d$}}
\put(72.50,77.00){\oval(45.00,16.00)[b]}
\put(50.00,80.00){\line(1,1){23.00}}
\put(30.00,80.00){\line(1,0){85.00}}
\put(30.00,79.00){\line(1,0){20.00}}
\put(30.00,81.00){\line(1,0){20.00}}
\put(95.00,80.00){\line(0,-1){1.00}}
\put(95.00,79.00){\line(1,0){20.00}}
\put(95.00,81.00){\line(1,0){20.00}}
\put(95.00,81.00){\line(-1,1){22.00}}
\put(72.15,103.00){\line(0,1){19.00}}
\put(73.55,103.00){\line(0,1){19.00}}
\end{picture}
\centerline{\bf FIG. I}
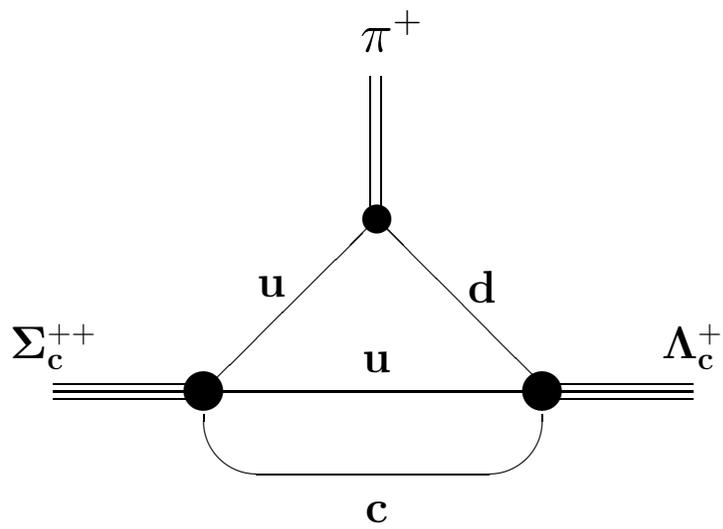
\end{document}